\documentclass[twocolumn]{article}
\usepackage{lmodern}
\usepackage{preprint}
\usepackage{cite}
\usepackage{amsmath, amsthm, amssymb, amsfonts}
\usepackage{graphicx}
\usepackage{textcomp}
\usepackage{multirow}
\usepackage{array}

\usepackage[utf8]{inputenc}	
\usepackage[T1]{fontenc}	
\usepackage{xcolor}		
\usepackage[colorlinks = true,
            linkcolor = purple,
            urlcolor  = blue,
            citecolor = cyan,
            anchorcolor = black]{hyperref}	
\PassOptionsToPackage{hyphens}{url}\usepackage{hyperref}
\usepackage{booktabs} 		
\usepackage{nicefrac}		
\usepackage{microtype}		
\usepackage{float}			
\usepackage[most]{tcolorbox}
\usepackage{subcaption}
\usepackage{tabularx}
\usepackage{comment}
\usepackage{xurl}
\newtcbox{\tok}{
  on line, enhanced, box align=base,
  colback=gray!14, colframe=black!22,
  boxrule=0.28pt, arc=1.0pt, auto outer arc,
  left=1pt, right=1pt, top=0pt, bottom=0pt, boxsep=0pt,
  nobeforeafter, tcbox raise base,
  fontupper=\sffamily\fontsize{8pt}{9.2pt}\selectfont
}

\usepackage{newfloat}
\DeclareFloatingEnvironment[name={Supplementary Figure}]{suppfigure}
\usepackage{sidecap}
\sidecaptionvpos{figure}{c}

\usepackage{titlesec}
\titlespacing\section{0pt}{6pt plus 2pt minus 2pt}{1pt plus 1pt minus 1pt} 
\titlespacing\subsection{0pt}{10pt plus 3pt minus 3pt}{1pt plus 1pt minus 1pt}
\titlespacing\subsubsection{0pt}{8pt plus 3pt minus 3pt}{1pt plus 1pt minus 1pt}
\renewcommand{\theparagraph}{\alph{paragraph}}
\titleformat{\paragraph}[runin]{\normalfont\normalsize\itshape}{\theparagraph)}{0.5em}{}[:]
\titlespacing{\paragraph}{0pt}{4pt plus 1pt minus 1pt}{0.5em}

\title{NoisePQC++: A Unified NIST-Compliant PQC and Hybrid-PQC Implementation of the Noise Protocol}

\author{
  Nadeem Ahmed \quad Aryya Gangopadhyay \quad Lei Zhang\\
  Department of Information Systems, University of Maryland Baltimore County, Baltimore, USA\\
  \texttt{\{nahmed5, gangopad, leizhang\}@umbc.edu}
}

\begin{document}
\pagestyle{plain}
\setlength{\footskip}{25pt}
\setlength{\parindent}{1.5em}
\setlength{\parskip}{0pt}

\twocolumn[ 
  \begin{@twocolumnfalse} 
  
\maketitle

\begin{abstract}
The threat of quantum computers to classical public-key cryptography has created an urgent need to evolve secure communication protocols with post-quantum cryptographic (PQC) primitives. The Noise Protocol Framework, widely used in systems such as WireGuard and WhatsApp, traditionally relies on the Elliptic Curve Diffie-Hellman (ECDH) public-key exchange scheme, which is vulnerable to quantum threats. In this paper, we present NoisePQC++, a unified C++23 implementation of the Noise Protocol framework augmented with post-quantum Key Encapsulation Mechanisms and Hybrid Forward Secrecy. Our design integrates the National Institute of Standards and Technology (NIST) standardized ML-KEM algorithm alongside classical ECDH, enabling full PQC, hybrid ECDH+PQC handshakes, and unified support for all 57 classical Noise handshake pattern variants, 13 post-quantum Noise handshakes, and their hybrid variants. Compared with prior work, NoisePQC++ offers broader protocol coverage, more complete implementation support, and greater flexibility. Our evaluation shows minimal overhead under normal network conditions and acceptable overhead in adverse cases, while significantly improving resistance against quantum adversaries. These results indicate that NIST-standardized post-quantum and hybrid Noise handshakes are practical and provide a credible basis for future deployment.
\end{abstract}
\vspace{0.3cm}
\vspace{0.25cm}
\noindent\rule{\linewidth}{0.4pt}\par\smallskip
{\footnotesize\itshape Accepted for publication at the IEEE International Conference on
Quantum Computing and Engineering (QCE~2026). \textcopyright~2026 IEEE\@. Personal use
of this material is permitted. Permission from IEEE must be obtained for all other uses,
in any current or future media, including reprinting/republishing this material for
advertising or promotional purposes, creating new collective works, for resale or
redistribution to servers or lists, or reuse of any copyrighted component of this work
in other works.}\par
\vspace{0.15cm}

  \end{@twocolumnfalse} 
] 


\section{Introduction}
\label{sec:introduction}

The Noise Protocol Framework\cite{NoiseProtocolFramework} provides a concise and flexible method for specifying Authenticated Key Exchange (AKE) protocols using Symmetric and Asymmetric Cryptography. Due to its simplicity, efficiency, and strong security foundation, the Noise Protocol Framework underpins the security of a variety of software systems, such as WireGuard, Lightning, and WhatsApp\cite{NoiseProtocolFramework}. The framework is often viewed as a lightweight, pattern-driven approach to build secure channels similar to Transport Layer Security (TLS) in application-specific deployments because it exposes a compact language to describe AKE and transport protection\cite{kobeissi2019noise}. Noise is also related to the Signal\cite{DoubleRatchet} family of secure messaging designs through the shared design lineage: while it is not itself a ratcheting protocol, the official Noise specification explicitly cites the Key Derivation Function (KDF) chains used in the Double Ratchet Algorithm as an influence on its design\cite{NoiseProtocolFramework}. 

This reliance on classical ECDH also creates a long-term risk: sufficiently powerful quantum computers could undermine algorithms such as X25519 and X448 by weakening the hardness of the underlying discrete logarithm problem\cite{NISTPQCSRC2017}. This has motivated efforts to extend Noise with \textbf{post-quantum cryptography (PQC)}, particularly \textbf{Key Encapsulation Mechanisms (KEMs)}, either replacing or augmenting ECDH\cite{zhang2020quantum,zhang2023making}. Prior work\cite{angel2022post,renckens2024evaluation,lepistoJmlepistoClatter2026,KatzenpostNyquist2025} has demonstrated that post-quantum and hybrid extensions of Noise are feasible, but existing implementations remain limited in pattern coverage, hybrid support, or overall completeness. A detailed comparison of these efforts is provided in Section \ref{subsec:related} and Table \ref{tab:capability-matrix}.

In this work, we bridge this gap and advance the existing Noise Protocol with five main \textbf{contributions}. First, we present \textbf{\textit{NoisePQC++}}, a C++23 implementation of the Noise Protocol Framework spanning classical, post-quantum, and hybrid handshakes within one unified architecture. Second, the implementation supports all 57 classical Noise handshake pattern variants, all 13 published PQNoise\cite{angel2022post} patterns, and systematic hybrid ECDH+PQC\cite{NoiseprotocolNoise_hfs_spec2026} variants, providing broader coverage than prior implementations. Third, our implementation integrates the National Institute of Standards and Technology (NIST) standardized ML-KEM\cite{nistfips203} algorithm at three security levels (ML-KEM-512, ML-KEM-768, and ML-KEM-1024), in conjunction with classical ECDH, within a unified handshake engine. Fourth, it provides practical engineering support for cross-platform builds, instrumentation, and application-level integration, as well as developer-oriented capabilities including parsing utilities, examples, and comprehensive testing infrastructure. Fifth, we evaluate the practicality of these designs through initial benchmarking, showing that post-quantum and hybrid Noise handshakes remain feasible with manageable overhead. To our knowledge, \textbf{\textit{NoisePQC++}} provides one of the most complete and flexible PQC-capable Noise Protocol reference implementations currently available.

The remainder of this paper reviews the background and related work in Section~\ref{sec:back-related}, presents the design, methodology, and architecture of \textbf{\textit{NoisePQC++}} in Sections~\ref{sec:designgoals},~\ref{sec:methodology} and~\ref{sec:architecture-implementation}, evaluates its correctness and performance in Section~\ref{sec:evaluation}, and concludes with future directions in Section~\ref{sec:conclusion}.

\section{Background and Related Work}
\label{sec:back-related}

\subsection{Noise Protocol Framework and ECDH Handshakes}

Noise is a framework for constructing secure channel protocols from simple cryptographic building blocks. At its core, a Noise handshake pattern is a predefined sequence of message exchanges carrying public keys or ciphertexts\cite{NoiseSpec34}. The allowed tokens in Noise handshake patterns include:

\begin{itemize}
    \item \tok{e} \textit{\textbf{(Ephemeral DH public key)}}: A party generates a new ephemeral key pair and sends the public key.
    \item \tok{s} \textit{\textbf{(Static DH public key)}}: A party sends its static (long-term) public key, often after encrypting it under an ephemeral-derived key.
    \item \tok{ee}, \tok{es}, \tok{se}, \tok{ss} \textit{\textbf{(DH operations)}}: These tokens denote DH computations between the initiator's and responder's key shares, where the first letter refers to the initiator, the second to the responder, \tok{e} denotes ephemeral, and \tok{s} denotes static\cite{NoiseSpec34}. Thus, \tok{ee} combines both ephemeral keys, \tok{es} the initiator's ephemeral with the responder's static, \tok{se} the initiator's static with the responder's ephemeral, and \tok{ss} both static keys. Each DH result is mixed into the cryptographic state of the handshake.
\end{itemize}

In a Noise handshake, both parties maintain a \tok{SymmetricState} with two primary components: a \emph{\textbf{chaining key}} (usually denoted \tok{ck}), which accumulates shared secrets to derive session keys, and a \emph{\textbf{handshake hash}} (\tok{h}), which accumulates transcript data to provide context integrity, often referred to as \emph{\textbf{channel binding}}. As each token is processed, new data is mixed into these states. At the end of the handshake, both parties derive symmetric secrets (e.g., encryption keys) from the final \tok{ck} and \tok{h}. If an adversary has not compromised participant secrets, the resulting session keys provide confidentiality and integrity for subsequent transport messages\cite{NoiseSpec34}.

Using these tokens, Noise defines 15 base patterns, including interactive patterns such as \textbf{NN, NK, NX, KN, KK, KX, XN, XK, XX, IN, IK,} and \textbf{IX}, which cover different authentication scenarios\cite{NoiseSpec34}. With one-way and pre-shared-key (PSK) modifiers, the Noise specification defines 57 variants. In these pattern names, the first letter describes the initiator’s static key and the second the responder’s: \textbf{N} means no static key, \textbf{K} means the static key is already known to the peer, \textbf{X} means the static key is transmitted during the handshake, and \textbf{I} means the initiator sends its static key immediately in the first message, reducing identity hiding\cite{NoiseSpec34}. For example, the widely used \textbf{XX} pattern is a three-message handshake in which the initiator sends an ephemeral key, the responder replies with its ephemeral key and encrypted static public key, and the initiator then sends its encrypted static public key, with multiple ECDH operations mixed into the handshake state. All Noise patterns use ECDH as the only public-key primitive, commonly X25519 for 128-bit security or X448 for 224-bit security. The security of Noise handshakes is well studied under strong models, assuming the hardness of the DH problem\cite{sarkarOverviewDiscreteLogarithm2024}. The flexibility of the framework comes from the support of different patterns and cryptographic primitives, including ECDH curves, symmetric ciphers, and hash functions, while preserving a unified state-machine model and security properties.

\subsection{Post-Quantum Cryptography}

Post-quantum cryptography refers to cryptographic algorithms believed to be secure against quantum computer attacks\cite{NISTPQCSRC2017}. In the context of key exchange, the primary PQC primitive is the KEM\cite{nistfips203}. A KEM is a public-key encryption scheme optimized for deriving a shared secret. One party generates a random secret and encapsulates it to the other party’s public key, producing a ciphertext. The other party decapsulates this ciphertext with its private key to recover the secret. Both parties then share this secret, which can serve as a key for symmetric encryption. The crucial difference from ECDH is that with a KEM, only the sender performs public-key encryption (encapsulation) and the receiver performs a private-key operation (decapsulation), rather than both performing a symmetric operation as in ECDH\cite{NIST_FIPS203_2024}.

\textbf{CRYSTALS-Kyber}, standardized by NIST as \textbf{ML-KEM} (\textbf{M}odule-\textbf{L}attice \textbf{KEM}) in FIPS 203, is a lattice-based KEM whose security is based on the Module-LWE problem, which is believed to be resistant to quantum attacks\cite{nistfips203}. ML-KEM is available in parameter sets that offer security roughly equivalent to classical 128-bit, 192-bit, and 256-bit levels, namely ML-KEM-512, ML-KEM-768, and ML-KEM-1024\cite{nistfips203}. In \textbf{\textit{NoisePQC++}}, we integrate all three parameter sets. Across the ML-KEM parameter sets collectively, public keys are approximately 800 to 1568 bytes, significantly larger than the elliptic curve keys (32 to 56 bytes) and the corresponding KEM encapsulation ciphertexts are about 768 to 1568 bytes. This size difference has implications for handshake message sizes and network performance. However, ML-KEM’s computational efficiency is high: on modern CPUs it can encapsulate/decapsulate in tens of microseconds, comparable to or even faster than some ECDH operations, meaning the computational overhead of using ML-KEM in a handshake is relatively low. We present these results later in Section \ref{sec:evaluation}.

\subsection{Hybrid Key Exchange}

Given the uncertainties in the long-term security of new PQC schemes, combining classical and post-quantum methods in hybrid modes is widely recommended by many organizations, such as the Internet Engineering Task Force (IETF) and NIST, during the transition to PQC\cite{ietf-tls-hybrid-design,NIST_SP800227_2025,rfc9370}. A \textbf{hybrid key exchange} means that the parties perform both a classical ECDH exchange \textbf{\textit{and}} a PQ KEM exchange, and then combine the two resulting shared secrets into one composite key. In this way, even if one of the schemes is later broken, the other (hopefully still secure) scheme ensures that the session remains protected. In \textbf{\textit{NoisePQC++}}, we implement hybrid handshakes as \textbf{Hybrid Forward Secrecy (HFS) patterns}, denoted by appending an ``\textbf{hfs}'' suffix to a base pattern. For example, \textbf{XXhfs} is the hybrid variant of \textbf{XX}, which performs both an X25519 ECDH and an ML-KEM encapsulation within the same handshake. Our implementation follows the approach of the specifications of the KEM-based Hybrid Forward Secrecy in Noise Protocol, as well as Renckens et al. (2024) and related work\cite{NoiseprotocolNoise_hfs_spec2026,renckens2024evaluation}. 

\subsection{Related Work}
\label{subsec:related}

Prior work established the foundations for post-quantum extensions of the Noise framework, but existing efforts remained limited in scope, implementation maturity, or feature coverage.

\subsubsection{Foundational PQ Noise Designs and Implementations}
Post-quantum Noise was first formalized by Yawning Angel et al.\ in \textit{Post-Quantum Noise}, which introduced a generic method for converting classical Noise patterns into PQ variants by replacing DH operations with KEM-based constructions\cite{angel2022post}. Their analysis established a comparable security notion under the IND-CCA-secure KEM assumptions, and they released an experimental prototype by extending the Go-based \textit{\textbf{nyquist}} library\cite{YawningAngelNyquist2026}. However, that implementation remained a proof of concept, focused primarily on pure PQ handshakes, and did not provide integrated hybrid ECDH+KEM support. Subsequent implementation work, most notably \textbf{\textit{Clatter}} in Rust, advanced this line by supporting PQNoise patterns as well as hybrid constructions\cite{lepistoJmlepistoClatter2026}. Clatter emphasized memory safety, correctness, and embedded suitability through pure Rust support, and followed Noise revision 34. At the same time, its design intentionally narrowed scope: it supports only X25519 for classical ECDH, omits X448 as well as deferred and fallback patterns, and does not implement the standard Noise pattern-string parser. Instead, handshakes are instantiated programmatically, which reduces flexibility for dynamic negotiation. Thus, while Clatter demonstrates a practical PQ-capable Noise library, it does not attempt full pattern or tooling coverage.

\subsubsection{Performance Studies and Related Hybrid Efforts}
Renckens et al. further examined the practicality of PQ and hybrid Noise by extending the \textit{\textbf{noise-c}} reference implementation with ML-KEM-512 and hybrid variants of the fundamental interactive Noise patterns\cite{renckens2024evaluation}. Their experiments on laptop and embedded ARM platforms showed that under typical network conditions, the latency differences between classical Noise and PQNoise are small because network delay dominates cryptographic cost. Overhead became more visible under elevated packet loss, where the larger PQ public keys and ciphertexts increased retransmission costs. They also found that hybrid X25519+ML-KEM exchanges impose only modest additional cost when no extra round trips are introduced. Although this work demonstrated feasibility, the resulting code remained research-oriented, limited to a single KEM and a restricted subset of patterns. Related efforts include experimental post-quantum extensions of WireGuard\cite{hulsing2021post}, effectively applying hybrid concepts to the Noise IK pattern, as well as Katzenpost's HPN library\cite{KatzenpostNyquist2025}, which explored combining multiple KEMs within one design. These efforts further illustrate practical and research interest in hybrid Noise-style protocols, but they likewise do not provide a broad, unified, developer-oriented implementation of the full Noise design space.

\subsubsection{Summary of Gaps}

Taken together, prior work demonstrated that PQ and hybrid Noise are both theoretically sound and practically viable. However, no earlier implementation combined broad Noise pattern coverage with the engineering features needed to move from a research prototype to a reusable software base: a unified classical/PQ/hybrid execution model, portability across platforms, and developer-focused capabilities such as parsing, instrumentation, and extensive testing. This makes \textbf{\textit{NoisePQC++}} more practical than prior efforts because the same codebase can be built, integrated, debugged, and evaluated across different environments without redesigning the protocol logic for each variant. It also makes the system useful as a research platform, since new patterns, KEM choices, fallback mechanisms, and application integrations can be studied within one consistent implementation rather than across disconnected prototypes.

\section{Design Goals and System Overview}
\label{sec:designgoals}

The development of the \textbf{\textit{NoisePQC++}} system design was driven by a set of fundamental objectives:

\subsection{Unified Execution Model}
 We decided early on that the same core state machine should handle classical, post-quantum, and hybrid handshakes. In contrast to earlier approaches, instead of forking the codebase or implementing separate handshake routines for PQ patterns, we augment the Noise handshake logic with clear, well-defined extensions for PQ. This ensures consistency and avoids redundant code. Concretely, we treat ECDH and KEM as two facets of a generic “key exchange” interface, allowing them to plug into the handshake state machine interchangeably. The handshake pattern definition format was extended to include the new token types (\tok{ekem}, \tok{skem}, discussed in Section~\ref{sec:architecture-implementation}), but the parsing and execution engine remains unified.
 
\subsection{Leverage of Existing Crypto Primitives}
 Implementing cryptography correctly is difficult. We rely on the \textbf{Botan} cryptographic library\cite{BotanBotana} as our underlying crypto provider for all operations: symmetric ciphers, hash functions, ECDH, and KEM. Botan is a well-established C++ library with constant-time implementations and has already included NIST-standardized PQC algorithms. The choice of using Botan as our cryptographic backend is derived from our previous work, which surveyed the main open-source cryptographic libraries for their support of PQC\cite{11250135}.  By using Botan, we offload low-level crypto details (e.g., field arithmetic for X25519 or lattice math for ML-KEM) to a vetted source. This allowed us to focus on the protocol integration logic. It also simplifies cross-platform support, as Botan contains optimized routines for various architectures (using AES-NI, AVX2, etc. when available)\cite{BotanBotana}. Our build system includes only the absolute minimal cryptographic primitive modules from the latest stable Botan 3 release for all target platforms (Windows x64, Linux x64/ARM, macOS x64/ARM) \cite{ahmed2026noisepqcpp} and, once compiled as its own static library by the build system, we invoke Botan via its C++ API inside \textbf{\textit{NoisePQC++}}.
 
\subsection{Type Safety and Modularity}
 We adopted modern C++23 features (especially \textit{modules} and strong type \textit{enums}) to create a clear separation of concerns in the code. Each higher-level \textbf{\textit{NoisePQC++}} cryptographic primitive is encapsulated in its own module (\tok{Cipher}, \tok{Hash}, \tok{Dh/KEM}), and the handshake state machine operates on abstracted types (e.g., \tok{KeyPair} or \tok{SharedSecret}) without needing to know the algorithm details. We defined \texttt{enum} classes for algorithm selections (\tok{KeyAlgo}, \tok{CipherAlgo}, \tok{HashAlgo}) so that an invalid combination (like using a KEM where ECDH is required, or an unsupported algorithm name) can be caught at compile-time or initialization-time. Using C++ modules instead of traditional headers helps with encapsulation: internal implementation details of, say, ML-KEM key sizes are hidden behind the \tok{Dh} module, exposing only clean interfaces to the rest of the system.
 
\subsection{Minimal Intrusion for PQ/HFS}
 When adding PQ and HFS features, we aim to minimize changes to the established Noise protocol flow. The Noise symmetric state (the running handshake hash and chaining key) is used exactly as before; we simply mix in KEM-derived secrets using the same HMAC-based KDF (HKDF) \tok{MixKey} operations that mix ECDH outputs. The order of mixing follows the pattern token order, so the transcript hash and chaining key evolve in a canonical way, whether a secret came from an ECDH or a KEM. For example, in a hybrid handshake, if the pattern dictates an \tok{ee} (ECDH) followed by an \tok{ekem1} (KEM token, discussed in Section~\ref{sec:architecture-implementation}) in the same message, we call \tok{SymmetricState.MixKey(DH\_shared)} then \tok{MixKey(KEM\_shared)} in that sequence. This ensures that the final derived keys are a function of \textbf{\textit{both}} secrets. We did not need to invent a new key combination function: the Noise HKDF chaining already serves to “mix” multiple inputs serially, which effectively concatenates the entropy. This approach aligns with the recommendations for hybrid key combination (often an XOR or KDF combination) and has the advantage of staying within the proven design of Noise\cite{NIST_SP800227_2025,NoiseSpec34}.
 
\subsection{Instrumentation and Testing as First-Class Goals}
 From the outset, we intended to conduct thorough testing and to offer clear insight into how the protocol operates internally. We generated unit tests for each module and each variant of the handshake pattern (more than 200 tests with 500+ assertions in total) to verify the correctness. For classical patterns, we cross-checked against official Noise test vectors where available\cite{noiseprotocolTestVectors}. For PQ patterns, we validated that initiator and responder instances of the same pattern produce matching keys and that known-edge cases (like the wrong key or message reorder) trigger the expected errors. We also developed an instrumentation system, described later in Section \ref{sec:architecture-implementation}, that can log handshake progress step by step. This dual focus on testing and logging influenced our methodology: for instance, we wrote the handshake execution logic in a way that could easily hook into logging callbacks at each token processing step, and we maintained a debug mode that parallels the release logic to ensure any divergence would be caught by tests.

\subsection{Summary}
Using these objectives as our core guiding principles, we implemented \textbf{\textit{NoisePQC++}} in an iterative manner: first by implementing the classical Noise patterns, then adding KEM support in the \tok{Dh} module, defining PQ patterns, and finally adding hybrid patterns once the PQ pieces were stable. At each stage, tests were used to confirm that adding new functionality did not break existing behavior (e.g., classical pattern tests continued to pass when PQ code was added, thanks to careful isolation of code paths when KEMs are not in use).

\section{Methodology}
\label{sec:methodology}

This section outlines the methodology underlying the post-quantum and hybrid extensions of \textbf{\textit{NoisePQC++}}. We first describe how classical Noise handshake operations are translated into KEM-based post-quantum patterns, and then show how classical ECDH and ML-KEM are combined to achieve hybrid forward secrecy. Together, these choices motivate the architecture and implementation presented in the following section.

\subsection{KEM-Based Methodology for PQ Noise Patterns}

Integrating KEMs for pure PQ patterns into the Noise framework is non-trivial and requires defining new handshake tokens analogous to \tok{e} and \tok{ee}. The PQNoise proposal \cite{angel2022post} introduced two new tokens:

\begin{itemize}
    \item \tok{ekem} \textbf{\textit{(Encapsulate to recipient’s key)}}: This token indicates that the sender will perform a KEM encapsulation to either the receiver’s static or ephemeral public key (depending on the pattern context), send the resulting ciphertext, and both parties will derive a shared secret from it.
    \item \tok{skem} \textbf{\textit{(Static key encapsulation)}}: This token indicates that the sender will send a KEM ciphertext that encapsulates to the receiver’s static public key. In practice, \tok{skem} is used in patterns where one party’s static key needs PQ protection, analogous to how \tok{ss} covers static-static ECDH exchanges.
\end{itemize}

By substituting ECDH tokens with these KEM tokens, one can derive \textbf{PQ Noise patterns}. For example, the classical \textbf{XX} pattern involves tokens: \tok{e $\to$ e, ee, s $\to$ s, se}. The corresponding \textbf{pqXX} pattern replaces ECDH operations with KEM operations: an initiator’s ephemeral KEM public key is sent (\tok{e} as usual), the responder encapsulates to it (\tok{ekem} in responder’s message), and static keys are sent protected by the KEM-derived secret rather than an ECDH secret. The result is that both sides arrive at a shared session key via the KEM instead of ECDH. Formal analysis has shown that these PQNoise patterns can achieve similar security goals as their ECDH counterparts\cite{angel2022post}.

\subsection{Hybrid Forward Secrecy Methodology}

\begin{table}[t]
\centering
\footnotesize
\renewcommand{\arraystretch}{1.03}
\setlength{\tabcolsep}{4pt}
\caption{Notation used in the \texttt{XXhfs} handshake transcript.}
\label{tab:xxhfs-notation}
\begin{tabular}{@{}p{0.28\columnwidth}p{0.66\columnwidth}@{}}
\toprule
\textbf{Symbol} & \textbf{Meaning} \\
\midrule
$(s, S)$ & Static keypair (X25519) \\
$(e, E)$ & Ephemeral DH keypair (X25519) \\
$(sk^{e1}, K^{e1})$ & Ephemeral KEM keypair (ML-KEM-768) \\
$ct$ & KEM ciphertext \\
$k_{pq}$ & KEM shared secret (post-quantum) \\
$k_{ee}, k_{es}, k_{se}$ & DH shared secrets (classical) \\
$ck$ & Chaining key (updated by MixKey) \\
$h$ & Handshake hash \\
$k$ & Symmetric cipher key (from MixKey) \\
$c_S$ & Encrypted static public key (AEAD) \\
$\text{MixKey}(ck,\cdot)$ & $\text{HKDF}(ck,\text{input}) \to (ck', k)$ \\
$\text{MixHash}(h,d)$ & $h \gets \text{HASH}(h \| d)$ \\
$\text{EncryptAndHash}$ & AEAD encrypt, then MixHash \\
$\text{DecryptAndHash}$ & MixHash, then AEAD decrypt \\
\bottomrule
\end{tabular}
\end{table}

Conceptually, a hybrid Noise pattern is an interleaving of a classical pattern with a PQ pattern. To support hybrid Noise patterns, two additional tokens are also defined\cite{NoiseprotocolNoise_hfs_spec2026}:

\begin{itemize}
    \item \tok{e1}: A secondary ephemeral key (specifically used for a KEM). This token indicates that one party generates an ephemeral KEM key pair (ML-KEM) and sends the public key as part of the handshake message.
    \item \tok{ekem1}: An encapsulation to the secondary ephemeral key. This token indicates that the other party performs KEM encapsulation using the peer’s ephemeral KEM public key (\tok{e1}), sending the ciphertext.
\end{itemize}

Internally, an HFS handshake such as \textbf{XXhfs} will execute all the classical tokens of \textbf{XX} (e.g., \tok{e, ee, s, se}) and the PQ tokens (\tok{e1, ekem1}) in a coordinated manner. For example, in \textbf{XXhfs}:
\begin{enumerate}
    \item Initiator sends classical ephemeral \tok{e} and a PQ ephemeral \tok{e1} in its first message.
    \item Responder, in reply, sends classical ephemeral \tok{e} and static \tok{s} as in \textbf{XX}, and encapsulates to the initiator’s PQ ephemeral (\tok{ekem1} token), sending the ML-KEM ciphertext. Now responder derives two secrets: one from ECDH \tok{ee} and one from PQ \tok{ekem1}.
    \item Initiator receives the responder’s message, performs the ECDH and decapsulates the PQ ciphertext with its ephemeral KEM private key, \textbf{obtaining the same two shared secrets}.
\end{enumerate}

Then both secrets are mixed into the chaining key of the handshake (usually concatenated or XORed through the HKDF mixing process) to produce the final session keys. The effect is that the session keys are secure unless both X25519 and ML-KEM are broken. We maintain the Noise pattern’s rule that all operations are folded into the key schedule in a defined order, preserving forward secrecy and authentication properties. As an example, we provide a detailed protocol transcript of \textbf{XXhfs} in Figure \ref{fig:xxhfs-twocol} (notational references to the protocol transcript are given in Table \ref{tab:xxhfs-notation}). All hybrid patterns in \textbf{\textit{NoisePQC++}} (documented fully in our \tok{Patterns} module source, which is introduced in Section \ref{sec:architecture-implementation}, and the documentation guide in our repository\cite{ahmed2026noisepqcpp}) are constructed to preserve the message flow and ordering constraints of the base patterns, ensuring that no additional round trips are needed beyond what the slower primitive might require (in practice, the hybrid patterns have the same number of messages as the greater of the classical or PQ pattern). This property is \textit{hybrid forward secrecy} to emphasize that forward secrecy holds even in a post-quantum context and against retroactive attacks on one of the algorithms. 

\begin{figure}[t]
\centering
\footnotesize
\renewcommand{\arraystretch}{1.2}
\begin{tabular}{@{}p{0.22\textwidth}@{\hspace{0.04\textwidth}}p{0.23\textwidth}@{}}
\toprule
\textbf{Initiator (I)} & \textbf{Responder (R)} \\
\midrule
\multicolumn{2}{@{}l@{}}{\textit{Key material:}} \\
Static: $(s_I, S_I)$ via X25519 & Static: $(s_R, S_R)$ via X25519 \\
\multicolumn{2}{@{}l@{}}{\textit{State:} $ck, h \gets \text{Initialize}(\texttt{protocol\_name})$} \\[4pt]
\midrule
\multicolumn{2}{@{}c@{}}{\textbf{Message 1:} $\boldsymbol{\to}$ \texttt{e, e1}} \\[2pt]
\textbf{Writes:} &
    \textbf{Reads:} \\
\texttt{e:} $(e_I, E_I) \gets \text{DH.Gen}()$ &
    Store $E_I$; \,$h \gets \text{MixHash}(h, E_I)$ \\
$h \gets \text{MixHash}(h, E_I)$ & \\
\texttt{e1:} $(sk_I^{e1}, K_I^{e1}) \gets \text{KEM.Gen}()$ &
    Store $K_I^{e1}$;\, $h \gets \text{MixHash}(h, K_I^{e1})$ \\
$h \gets \text{MixHash}(h, K_I^{e1})$ & \\
Send: $E_I \| K_I^{e1}$ & \\[4pt]
\midrule
\multicolumn{2}{@{}c@{}}{\textbf{Message 2:} $\boldsymbol{\leftarrow}$ \texttt{e, ee, ekem1, s, es}} \\[2pt]
\textbf{Reads:} &
    \textbf{Writes:} \\
& \texttt{e:} $(e_R, E_R) \gets \text{DH.Gen}()$ \\
Store $E_R$;\, $h \gets \text{MixHash}(h, E_R)$ &
    $h \gets \text{MixHash}(h, E_R)$ \\[2pt]
\texttt{ee:} $k_{ee} \gets \text{DH}(e_I, E_R)$ &
    \texttt{ee:} $k_{ee} \gets \text{DH}(e_R, E_I)$ \\
$ck, k \gets \text{MixKey}(ck, k_{ee})$ &
    $ck, k \gets \text{MixKey}(ck, k_{ee})$ \\[2pt]
\texttt{ekem1:} $k_{pq} \gets \text{Decaps}(sk_I^{e1}, ct)$ &
    \texttt{ekem1:} $(ct, k_{pq}) \gets \text{Encaps}(K_I^{e1})$ \\
$ck, k \gets \text{MixKey}(ck, k_{pq})$ &
    $ck, k \gets \text{MixKey}(ck, k_{pq})$ \\[2pt]
\texttt{s:} $S_R \gets \text{DecryptAndHash}(c_S)$ &
    \texttt{s:} $c_S \gets \text{EncryptAndHash}(S_R)$ \\[2pt]
\texttt{es:} $k_{es} \gets \text{DH}(e_I, S_R)$ &
    \texttt{es:} $k_{es} \gets \text{DH}(s_R, E_I)$ \\
$ck, k \gets \text{MixKey}(ck, k_{es})$ &
    $ck, k \gets \text{MixKey}(ck, k_{es})$ \\
& Send: $E_R \| ct \| c_S$ \\[4pt]
\midrule
\multicolumn{2}{@{}c@{}}{\textbf{Message 3:} $\boldsymbol{\to}$ \texttt{s, se}} \\[2pt]
\textbf{Writes:} &
    \textbf{Reads:} \\
\texttt{s:} $c_S \gets \text{EncryptAndHash}(S_I)$ &
    \texttt{s:} $S_I \gets \text{DecryptAndHash}(c_S)$ \\[2pt]
\texttt{se:} $k_{se} \gets \text{DH}(s_I, E_R)$ &
    \texttt{se:} $k_{se} \gets \text{DH}(e_R, S_I)$ \\
$ck, k \gets \text{MixKey}(ck, k_{se})$ &
    $ck, k \gets \text{MixKey}(ck, k_{se})$ \\
Send: $c_S$ & \\[4pt]
\midrule
\multicolumn{2}{@{}c@{}}{\textbf{Split:} $(k_{I \to R},\; k_{R \to I}) \gets \text{HKDF}(ck, \varnothing)$} \\
\multicolumn{2}{@{}c@{}}{$\Rightarrow$ Transport encryption with derived session keys} \\
\bottomrule
\end{tabular}
\caption{Two-party protocol transcript of the \texttt{XXhfs} handshake
    (\texttt{Noise\_XXhfs\_25519+M768\_ChaChaPoly\_SHA256}).}
\label{fig:xxhfs-twocol}
\end{figure}

\section{Architecture and Implementation}
\label{sec:architecture-implementation}

\subsection{Layered Module Architecture}

\begin{figure}[t]
    \centering
    \includegraphics
    [width=\linewidth,trim={6mm 4mm 6mm 4mm},clip]{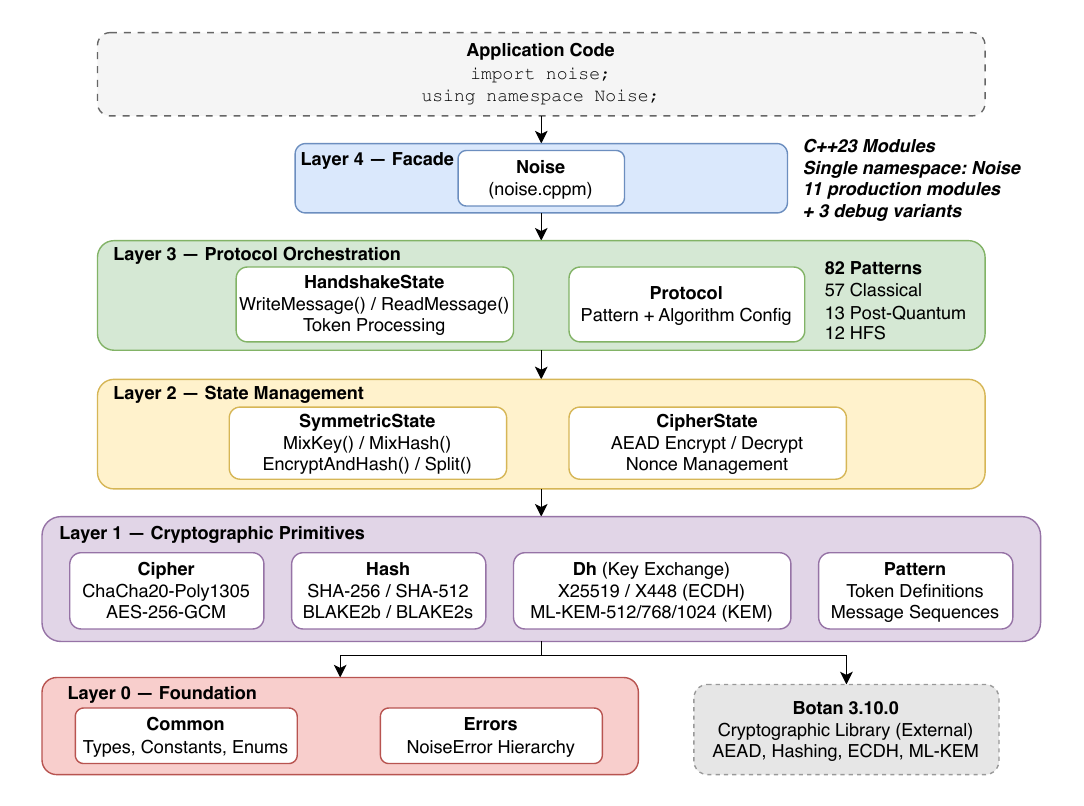}
    \caption{Shows the 5-layer module architecture of \textbf{\textit{NoisePQC++}}.}
    \label{fig:noisepqc-arch}
\end{figure}

\textbf{\textit{NoisePQC++}} is organized as layered C++ modules that mirror the structure of the Noise protocol. It separates cryptographic primitives, protocol state objects, and handshake logic. Figure \ref{fig:noisepqc-arch} shows the overall architecture.

\subsubsection{\textbf{Layer 0}}
Layer 0 provides shared utilities and error handling. The \tok{Common} module defines core types, constants, algorithm identifiers, compile-time options, and logging controls. The \tok{Errors} module defines the exception hierarchy, including \tok{NoiseError}, \tok{HandshakeError}, and \tok{InvalidTokenError}.

\subsubsection{\textbf{Layer 1}}
Layer 1 contains the cryptographic primitive modules. Except for the KEM algorithm, all other algorithms supported in this layer are based on the Noise Protocol Framework Specifications\cite{NoiseSpec34}.

\paragraph{\textbf{Cipher}}
Wraps AEAD encryption and decryption. We support \textbf{ChaCha20-Poly1305} and \textbf{AES-256-GCM} through Botan. Nonce management is handled primarily by \tok{CipherState}, and key material is wiped when no longer needed.

\paragraph{\textbf{Hash}}
Provides hashing, HMAC, and HKDF functionality. Supported algorithms are \textbf{SHA-256}, \textbf{SHA-512}, \textbf{BLAKE2b}, and \textbf{BLAKE2s}. The module supports both one-shot and incremental hashing for transcript maintenance and key derivation.

\paragraph{\textbf{Dh}}
Unifies classical ECDH and post-quantum KEM operations behind one interface. It supports \textbf{X25519} and \textbf{X448} for ECDH, and \textbf{ML-KEM-512/768/1024} for KEM. The interface exposes key generation, ECDH shared-secret derivation, and KEM encapsulation/decapsulation. A \tok{KeyAlgo} enum identifies the active algorithm, and higher layers use \tok{IsKEM()} to select ECDH or KEM logic. Each algorithm produces its shared secret at its specified length (32 bytes for X25519 and all ML-KEM parameter sets, 56 bytes for X448); every secret is absorbed through the same HKDF-based \tok{MixKey()} operation, so downstream code treats ECDH and KEM outputs uniformly.

\paragraph{\textbf{Pattern}}
Represents handshake patterns as token sequences grouped by message. \textbf{\textit{NoisePQC++}} includes standard classical Noise patterns, published PQNoise patterns, and HFS variants. The module also validates token ordering and related constraints before execution.

\subsubsection{\textbf{Layer 2}}
Layer 2 manages the evolving protocol state.

\paragraph{\textbf{CipherState}}
Tracks the key and nonce for one AEAD instance. After the handshake, two \tok{CipherState} objects are used for transport encryption, one per direction. The module provides authenticated encryption/decryption, automatic nonce incrementing, optional \tok{Rekey()} support, and overflow checks.

\paragraph{\textbf{SymmetricState}}
Maintains the handshake’s \textbf{chaining key \tok{(ck)}} and \textbf{handshake hash \tok{(h)}}. It is initialized from the protocol name and provides \tok{MixKey()}, \tok{MixHash()}, and \tok{Split()}. ECDH and KEM outputs are handled identically by mixing the resulting shared secret into \tok{ck}.

\subsubsection{\textbf{Layer 3}}
Contains protocol and token processing logic.

\paragraph{\textbf{Protocol}}
Combines a pattern with concrete algorithm choices to form a full protocol configuration. Examples include \tok{Noise\_XX\_25519\_ChaChaPoly\_SHA256}, \tok{Noise\_pqXX\_M768\_ChaChaPoly\_SHA256}, and \tok{Noise\_XXhfs\_25519+M768\_ChaChaPoly\_SHA256}. The \tok{Protocol} class can be built from components or parsed from a string and enforces consistency between pattern type and selected algorithms.

\paragraph{\textbf{HandshakeState}}
Implements handshake execution and token processing. A \tok{HandshakeState} is initialized with a \tok{Protocol}, the required static keys, and the optional PSK or prologue data, then advances through the pattern using \tok{WriteMessage()} and \tok{ReadMessage()}. Internally, it maintains the \tok{SymmetricState}, handshake cipher material, the relevant local and remote ECDH/KEM keys, and pattern progress state. During execution, \tok{e} and \tok{e1} generate ephemeral public keys, \tok{s} sends the local static public key, \tok{ee}/\tok{es}/\tok{se}/\tok{ss} compute and mix ECDH shared secrets, \tok{ekem}/\tok{ekem1}/\tok{skem} perform KEM encapsulation or decapsulation and mix the resulting shared secret, and \tok{psk} mixes the pre-shared secret. Handshake payloads are then processed with \tok{EncryptAndHash()} or \tok{DecryptAndHash()} to preserve transcript binding and authenticated protection.

This same structure supports hybrid handshakes: if a message contains both classical and PQ tokens, \tok{HandshakeState} executes them in declared order and mixes each resulting secret into the chaining key in sequence. 

After the final handshake message, \tok{Split()} derives the two transport \tok{CipherState} objects and the handshake transitions to transport mode. Any attempt to continue the handshake raises \tok{HandshakeDoneError}.

\subsubsection{\textbf{Layer 4}}
\begin{figure}[t]
    \centering
    \includegraphics[width=\linewidth,
    trim={10mm 12mm 10mm 12mm},
    clip]{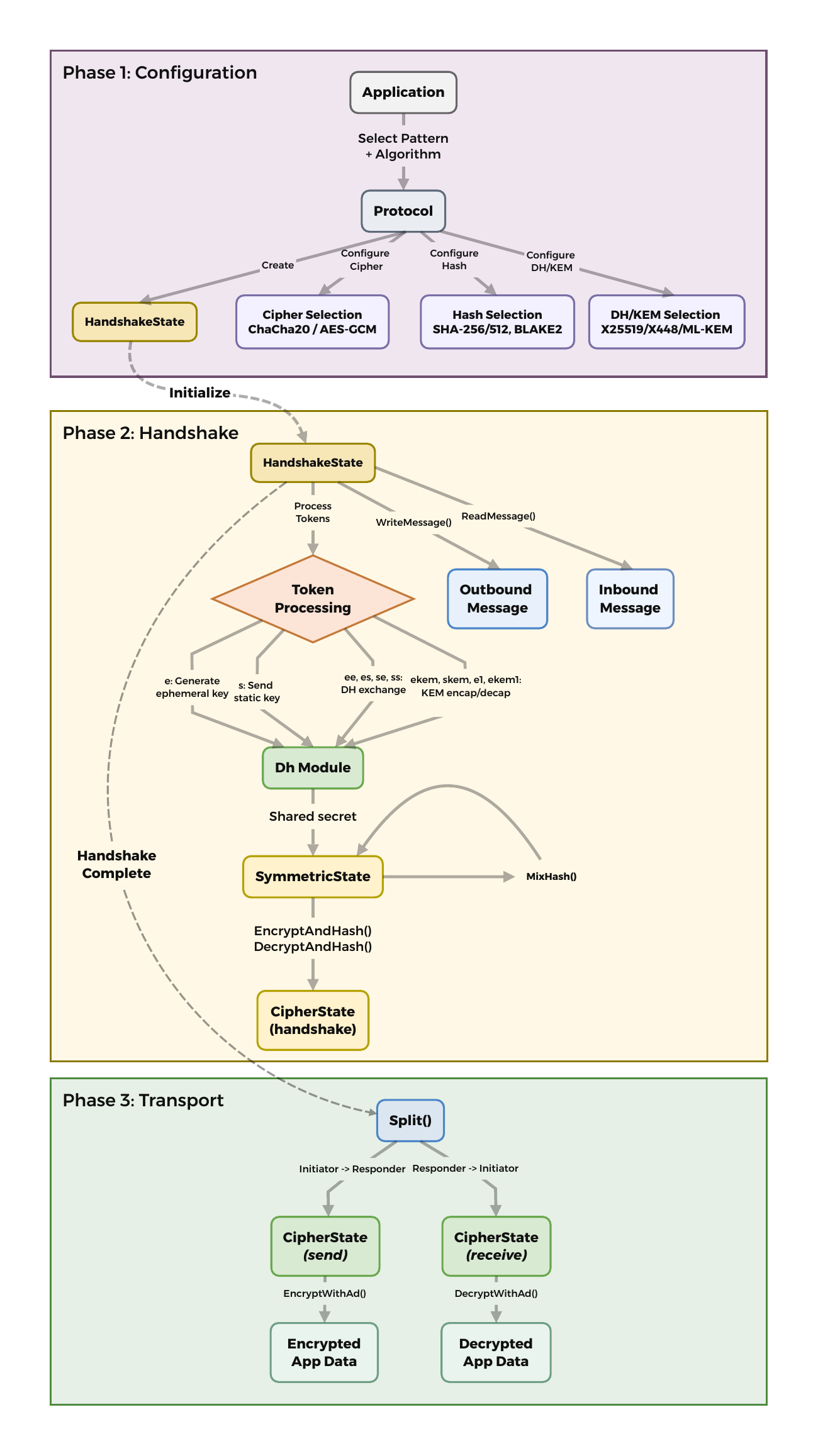}
    \caption{Runtime Handshake flow: app to transport encryption.}
    \label{fig:noisepqc-handshakeflow}
\end{figure}

Layer 4 exposes the \textbf{\textit{NoisePQC++ API}} to application code. Figure \ref{fig:noisepqc-handshakeflow} shows the runtime flow of handshakes from application calls to transport encryption.

\subsection{Implementation and Integration}

This section elaborates the practical engineering contributions introduced in Section~\ref{sec:introduction}: cross-platform build support, instrumentation, and integration with real networking environments.

\subsubsection{Cross-Platform Build and Botan Integration}

Building a C++23 codebase with modules and a large cryptographic dependency across platforms is non-trivial. We use \textbf{\textit{CMake}} (version $\geq$ 4.0) \cite{CMakeUpgradeYour} build system with the \textbf{\textit{Ninja}} generator \cite{NinjaSmallBuild} and require modern compilers with module support: \textbf{\textit{Clang 18.1}}\cite{Clang1816Release}, \textbf{\textit{GCC 15.2}}\cite{GCC15Release}, or \textbf{\textit{MSVC 19.44}}\cite{tylermsftVisualStudio}. Custom toolchain files and test scripts validate builds on Linux, macOS, and Windows. To simplify builds, we bundle the minimum required cryptographic source modules from the latest stable Botan 3 release for each supported platform and architecture, following its minimized build guide\cite{BuildingLibraryBotan}. \textit{CMake} selects the correct platform sources and compiles them into a standalone static library. This proved simpler for C++ module integration and improves portability by avoiding runtime library dependencies. Within the implementation, Botan is used through its standard APIs. For example, X25519 key generation uses Botan’s \tok{PK\_Key\_Agreement\_Key}, while ML-KEM encapsulation uses Botan’s KEM interfaces for the selected parameter set. Memory safety is supported through C++ RAII\cite{RAIICppreferencecom}, secure containers where possible, memory scrubbing on destruction, and spans and smart pointers instead of raw pointers. \textbf{\textit{NoisePQC++}} is currently single-threaded at the \tok{HandshakeState} level to reduce the complexity of protocol state management, with the potential to adopt a multithreaded architecture in future revisions.

\begin{table*}[!t]
\centering
\caption{Feature comparison of post-quantum Noise implementations}
\label{tab:capability-matrix}
\footnotesize
\setlength{\tabcolsep}{3pt}
\renewcommand{\arraystretch}{1.08}
\begin{tabular}{@{}>{\hspace{0pt}}m{0.27\textwidth}>{\centering\hspace{0pt}}m{0.14\textwidth}>{\centering\hspace{0pt}}m{0.15\textwidth}>{\centering\hspace{0pt}}m{0.14\textwidth}>{\centering\arraybackslash\hspace{0pt}}m{0.18\textwidth}@{}}
\toprule
\textbf{Capability / Feature} & \textbf{PQNoise} & \textbf{SecITC 2024 Proto} & \textbf{Clatter} & \textbf{NoisePQC++ (ours)} \\
\midrule
All 57 classical handshake variants & Yes (in base library) & Partial (12 core) & Partial (omits some) & \textbf{Yes (full)} \\
\midrule
PQNoise patterns (13) & Partial (some tested) & Yes (12 fundamental) & Yes (all 13) & \textbf{Yes (all 13)} \\
\midrule
Hybrid (classical+PQ) patterns & No & Yes (12 combined) & Yes (Hybrid) & \textbf{Yes (all interactive)} \\
\midrule
Multi KEM (ML-KEM-512/768/1024) & Possibly 512 only & 512 only & Yes (512,768,1024) & \textbf{Yes (512,768,1024)} \\
\midrule
DH support (25519, 448) & Yes & Yes & 25519 only & \textbf{Yes (25519, 448)} \\
\midrule
Static key authentication & Yes & Yes & Yes & \textbf{Yes} \\
\midrule
PSK support & Yes & Not mentioned & ? (not explicit) & \textbf{Yes (all psk patterns)} \\
\midrule
Deferred patterns & Yes (in base Noise) & Not implemented & No [12] & \textbf{Yes (all deferred)} \\
\midrule
Fallback patterns & No & No & No & \textbf{No (planned)}\\
\midrule
Implementation language & Go (prototype) & C (Noise-C modified) & Rust & \textbf{C++23} \\
\midrule
Logging/Instrumentation & Minimal & No & Minimal & \textbf{Extensive (color logs)} \\
\midrule
Test suite & Basic sanity tests & Experiment scripts & Yes (good) & \textbf{Comprehensive }\\
\midrule
Platform support & Linux/Unix (Go) & Linux (targeted) & All (Rust) & \textbf{Win/Linux/macOS }\\
\bottomrule
\end{tabular}
\end{table*}

\subsubsection{Instrumentation and Logging}

A distinguishing feature of \textbf{\textit{NoisePQC++}} is its handshake instrumentation for debugging and analysis. In debug builds, the library can log token processing, state transitions, key-mixing events, and other internal protocol steps. Rather than scattering statements of \tok{\#ifdef DEBUG} throughout production code, we maintain parallel debug variants of core state-machine modules, including \tok{HandshakeState.debug} and \tok{SymmetricState.debug}. A \textit{CMake} option selects either the production or debug implementation at compile time. As a result, production builds incur zero logging overhead because the debug code is never compiled into release binaries.

By default, debug builds enable full logging. The output is human-readable and color-coded, with categories for state transitions, token handling, and cryptographic events. To avoid accidental disclosure, logging is by default conservative: it reports metadata such as sizes, roles, and protocol events, but not secret key material. These categories are documented in the project’s Logging guide, and users can disable logging globally or selectively through API calls\cite{ahmed2026noisepqcpp}. This instrumentation is especially valuable for experimental and research use. Users can run example handshakes, including the interactive “pattern zoo,” and observe how the transcript state evolves and how hybrid handshakes combine multiple secrets. This transforms the protocol from an opaque “black-box” process into a transparent and analyzable execution trace and, to the best of our knowledge, is not a common feature in existing Noise protocol implementations. Complete color coded demo debug logs of \textbf{XX}, \textbf{pqXX} and \textbf{XXhfs} are available in our repository under the \texttt{output} folder\cite{ahmed2026noisepqcpp}.

\subsubsection{Example Integration and Use Cases}

To demonstrate deployability, we developed several example applications around \textbf{\textit{NoisePQC++}}. One is a simple ASIO-based\cite{AsioLibrary} client-server pair that performs a Noise handshake over TCP. We also provide a multi-protocol Noise server that accepts classical \textbf{XX}, \textbf{pqXX}, or hybrid \textbf{XXhfs} handshakes. The server identifies the handshake variant from the initial message format and instantiates the corresponding \tok{HandshakeState}. Because classical, PQ, and hybrid modes share the same library interface, the surrounding application logic remains nearly identical; in most cases, only the protocol name string changes. We also include an interactive Pattern Explorer that lets users select any classical, PQ, or hybrid pattern and observe a simulated handshake through the logging system. For example, a \textbf{pqXX} execution visibly includes an additional KEM ciphertext and corresponding \tok{ekem} handling compared to classical \textbf{XX}.

For validation, we integrate the official Noise test vectors for classical patterns\cite{noiseprotocolTestVectors}. Since there is no standardized test-vector suite for PQNoise or hybrid patterns, we generated deterministic test cases using fixed random seeds. These provide reproducible outputs and can serve as reference vectors for future implementations. \textbf{\textit{NoisePQC++}} provides a single public module, \tok{Noise}, which exposes core abstractions such as \tok{Protocol}, \tok{HandshakeState}, and \tok{CipherState}. These abstractions constitute the primary interface through which higher-level systems and applications can be systematically integrated with the library. Botan remains an internal dependency and does not leak into the public API, lowering the barrier to adopting a post-quantum-capable implementation. At the same time, production deployment of PQ handshakes should remain cautious. Although ML-KEM has been standardized, post-quantum algorithms are newer and less operationally mature than classical algorithms. Hence, the hybrid mode is often the most prudent default. \textbf{\textit{NoisePQC++}} supports this directly through \tok{hfs} patterns and protocol strings such as \tok{25519+M768}, enabling early experimentation with quantum-resistant secure channels in applications such as VPNs and messaging systems.

\section{Evaluation}
\label{sec:evaluation}

We evaluate \textbf{\textit{NoisePQC++}} along three dimensions: feature coverage, correctness, and performance. First, we compare its protocol support and flexibility with prior implementations. Second, we assess the correctness through testing and expected interoperability behavior. Third, we measure the practical overhead of classical, post-quantum, and hybrid handshakes across latency, CPU cost, message size, and network performance.

\subsection{Feature Coverage and Correctness}

Table \ref{tab:capability-matrix} provides a high-level comparison of \textbf{\textit{NoisePQC++}} with the prior implementations discussed (PQNoise, SecITC prototype, Clatter)\cite{angel2022post,renckens2024evaluation,lepistoJmlepistoClatter2026}. As shown, \textbf{\textit{NoisePQC++}} offers the broadest support for patterns and algorithms. Importantly, it is the only one that we know of that covers the \textbf{deferred} pattern variants (those with a ``1'' token, which are edge cases for when static keys are sent later in the handshake) and the various PSK patterns. Although these patterns are less commonly used, implementing them in the PQ and hybrid context demonstrates the generality of our approach. Clatter, for example, left deferred patterns to be “implemented by the user” if needed, whereas we included them by systematically extending the pattern generation rules. We also include both widely used curves (X25519, X448) and the full range of ML-KEM parameters, whereas others often fixed on a single parameter set for simplicity. This gives \textbf{\textit{NoisePQC++}} the flexibility to target different security levels (e.g., one could instantiate a Noise handshake with X448 + ML-KEM-1024 for a very high security margin, albeit with some performance cost). 

\subsection{Correctness and Interoperability}
Our test suite (we use \textbf{\textit{Catch2}}\cite{CatchorgCatch22026}, a popular C++ testing framework) confirms that for every pattern variant, the initiator and responder derive identical session keys and that security properties hold (e.g., if a wrong pre-shared key is used, the handshake fails with an error, as expected). We have not yet conducted formal interoperability testing with other libraries. However, by adhering closely to the specifications and using standardized algorithms, we expect that an initiator using \textbf{\textit{NoisePQC++}} and a responder using another library with the same pattern and algorithms will interoperate. This is an area for future empirical testing.

\begin{figure*}[!t]
\centering
\subfloat[Handshake latency comparison\label{fig:latency}]{
    \includegraphics[width=0.47\textwidth,trim=6 6 6 6,clip]{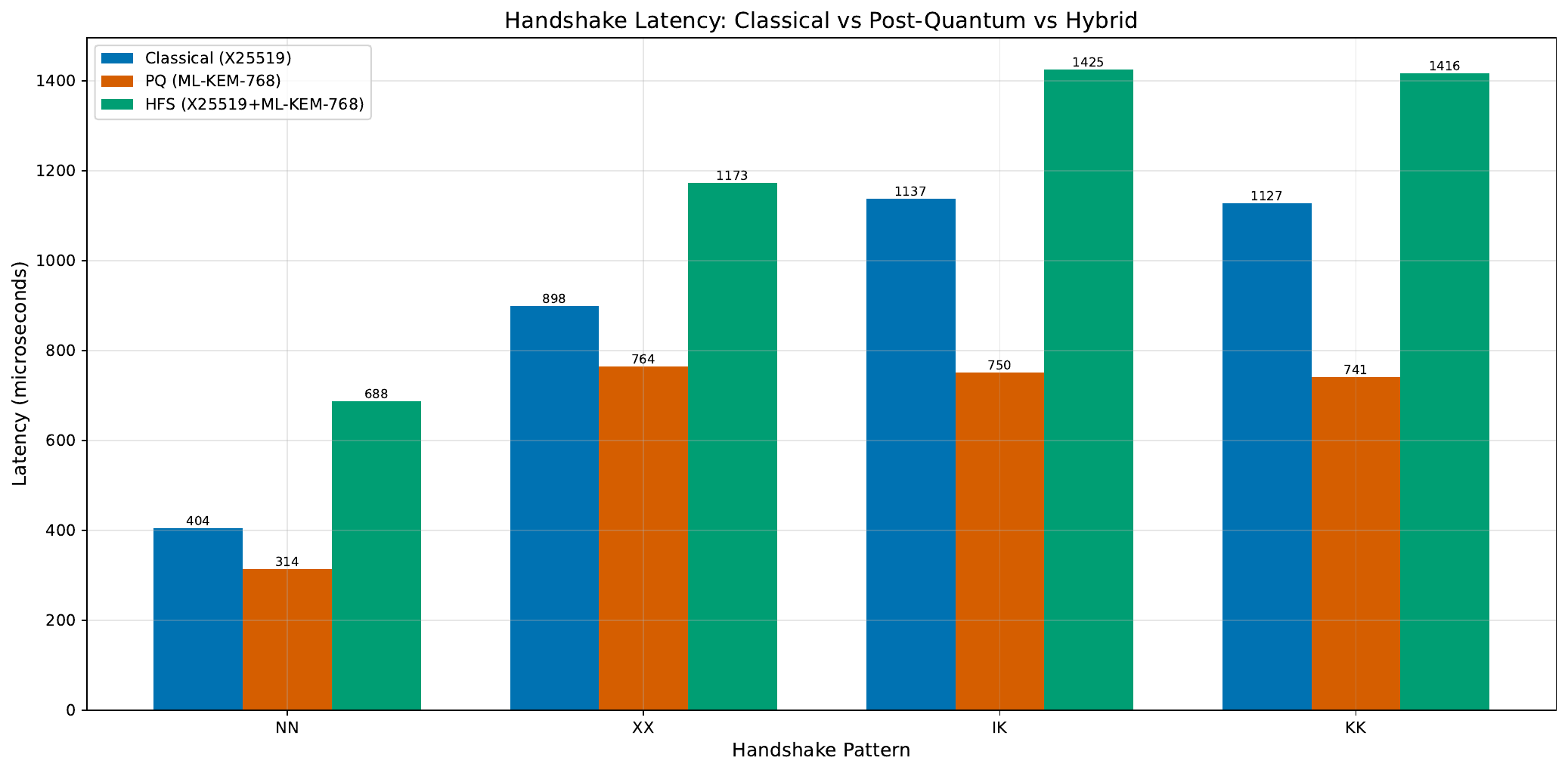}}
\hfill
\subfloat[Key generation performance\label{fig:keygen}]{
    \includegraphics[width=0.47\textwidth,trim=6 6 6 6,clip]{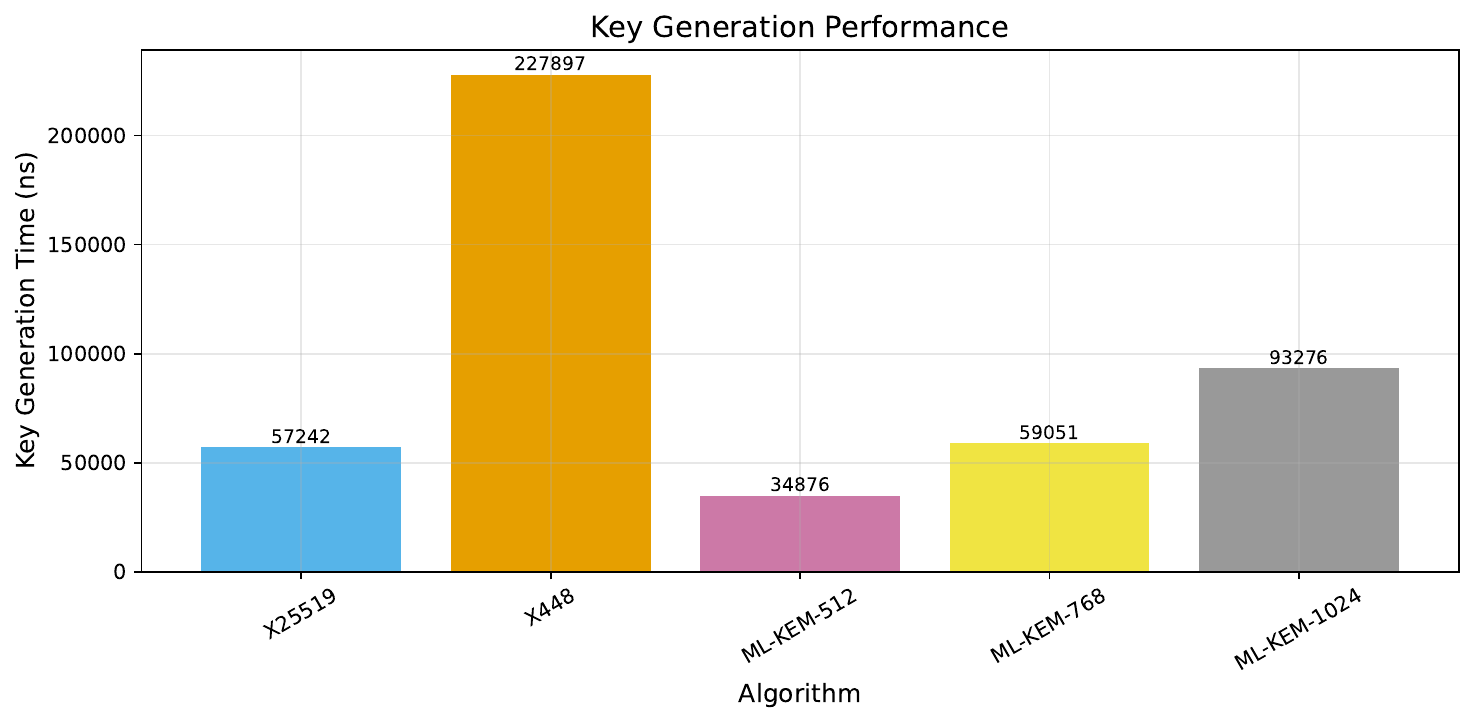}}

\vspace{2pt}

\subfloat[Message overhead per handshake\label{fig:overhead}]{
    \includegraphics[width=0.47\textwidth,trim=6 6 6 6,clip]{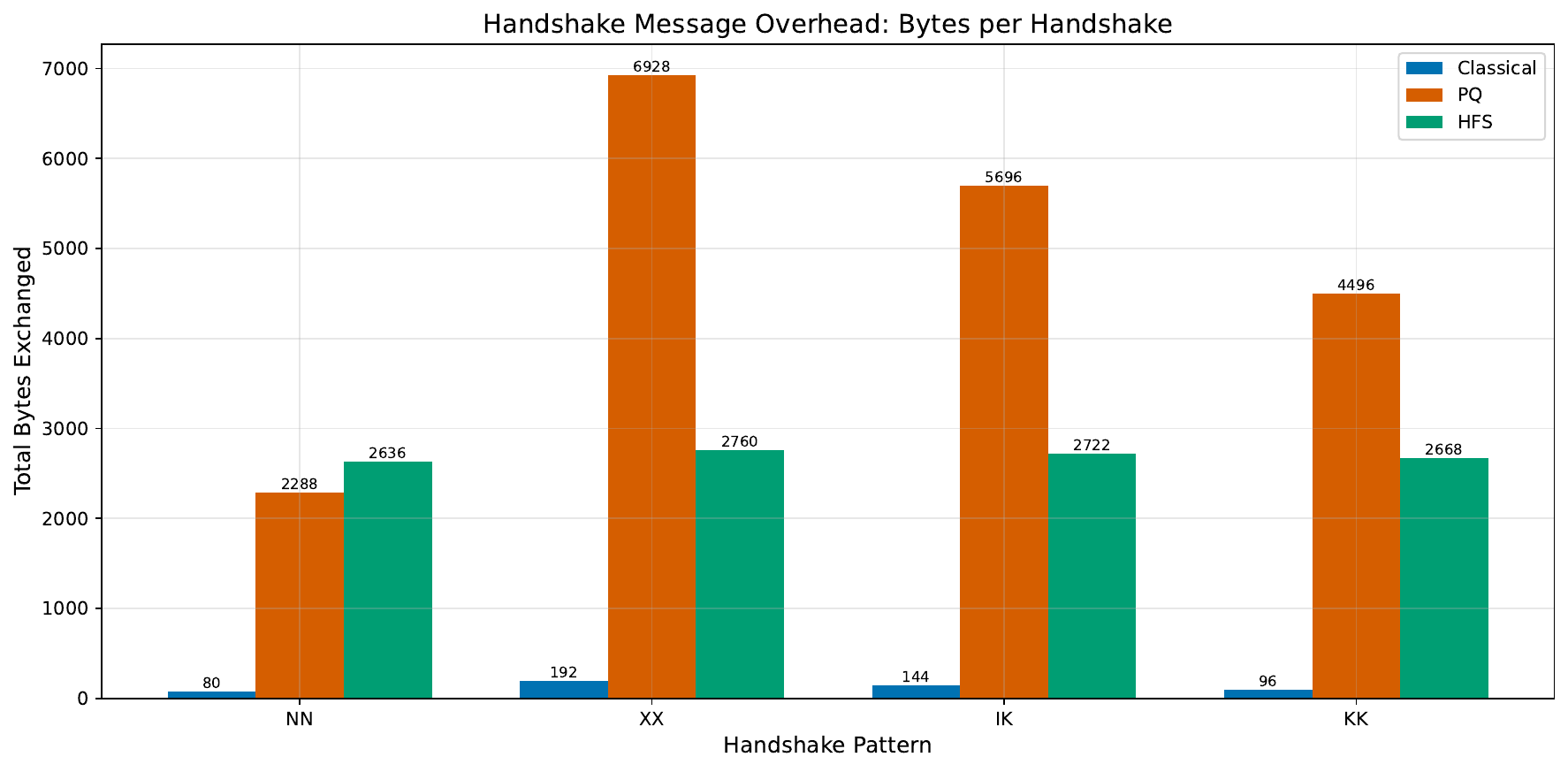}}
\hfill
\subfloat[Network round-trip time\label{fig:rtt}]{
    \includegraphics[width=0.47\textwidth,trim=6 6 6 6,clip]{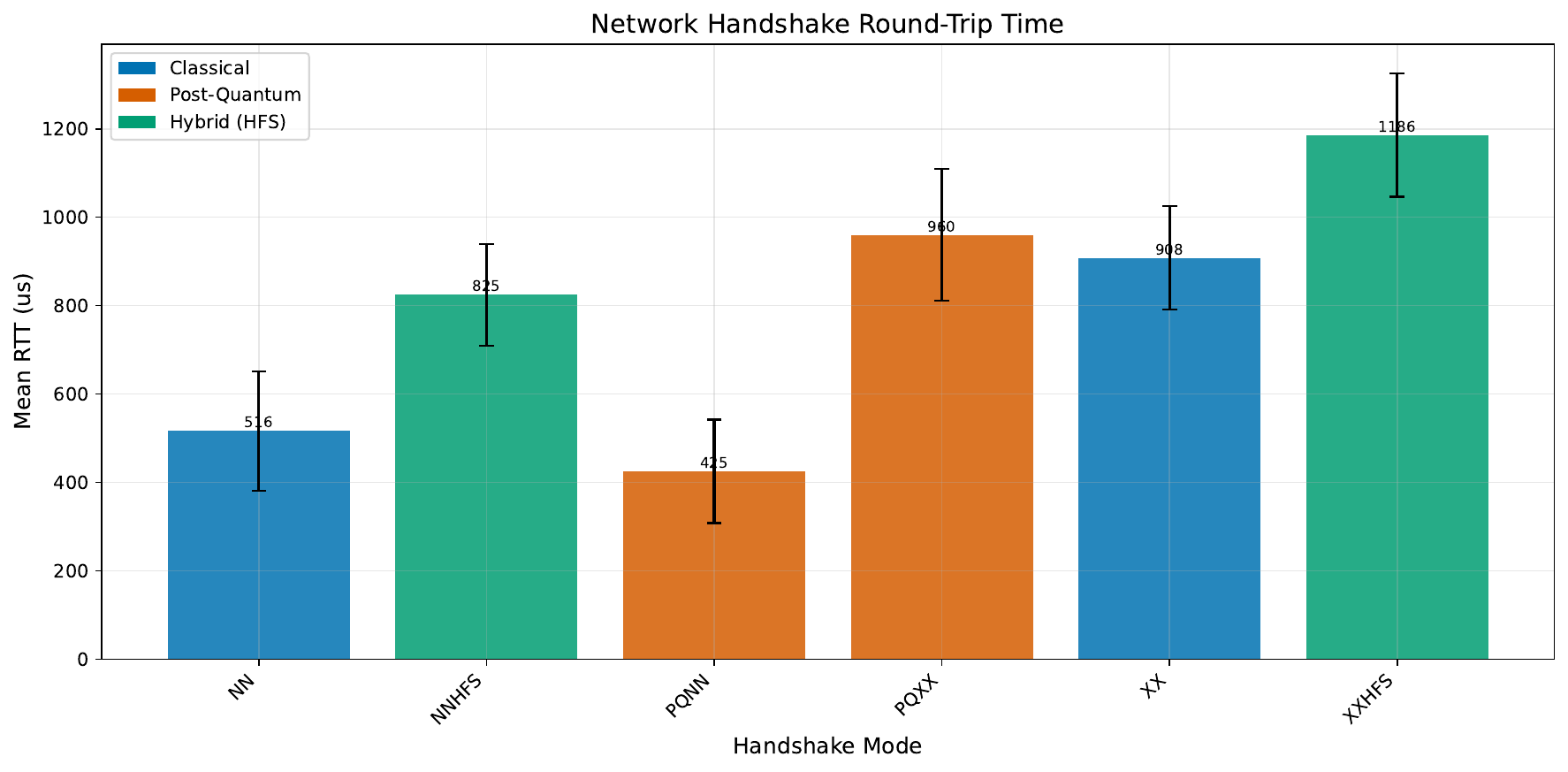}}

\caption{Benchmark results in classical (X25519), PQ (ML-KEM-768), and HFS (X25519+ML-KEM-768) modes.}
\label{fig:benchmarks}
\end{figure*}

\begin{figure}[!t]
\centering
\includegraphics[width=\columnwidth,trim=6 6 6 6,clip]{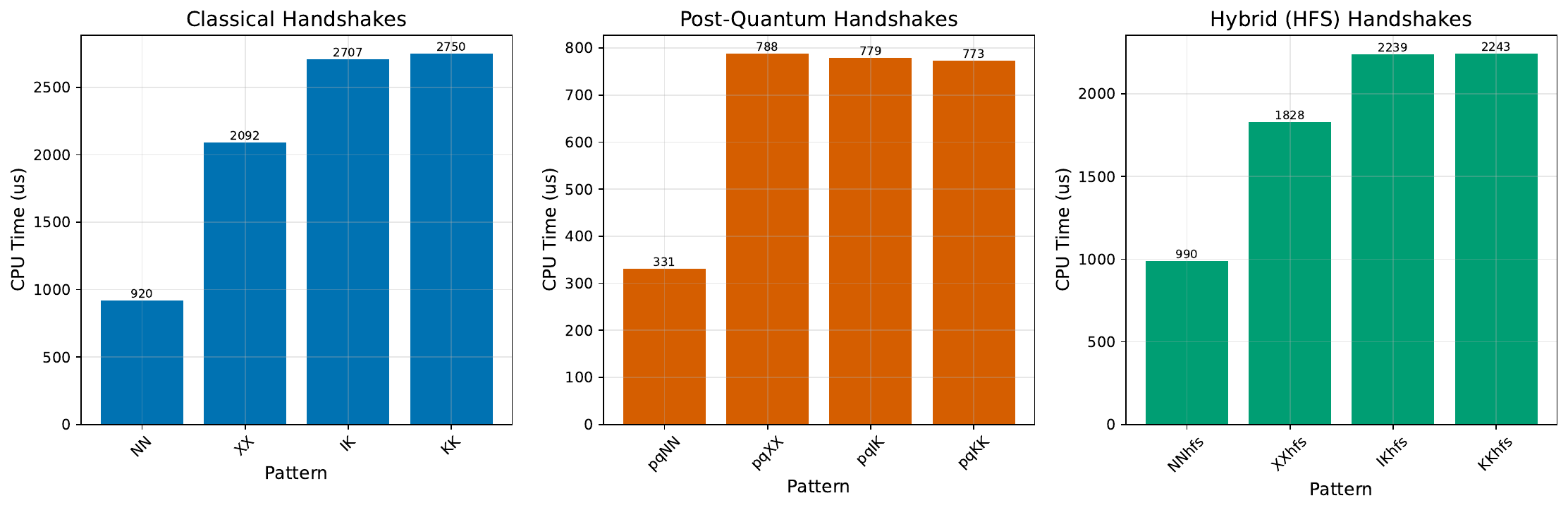}
\caption{CPU time across classical, PQ, and hybrid patterns.}
\label{fig:cpu}
\end{figure}

\begin{figure}[!t]
\centering
\includegraphics[width=\columnwidth,trim=6 6 6 6,clip]{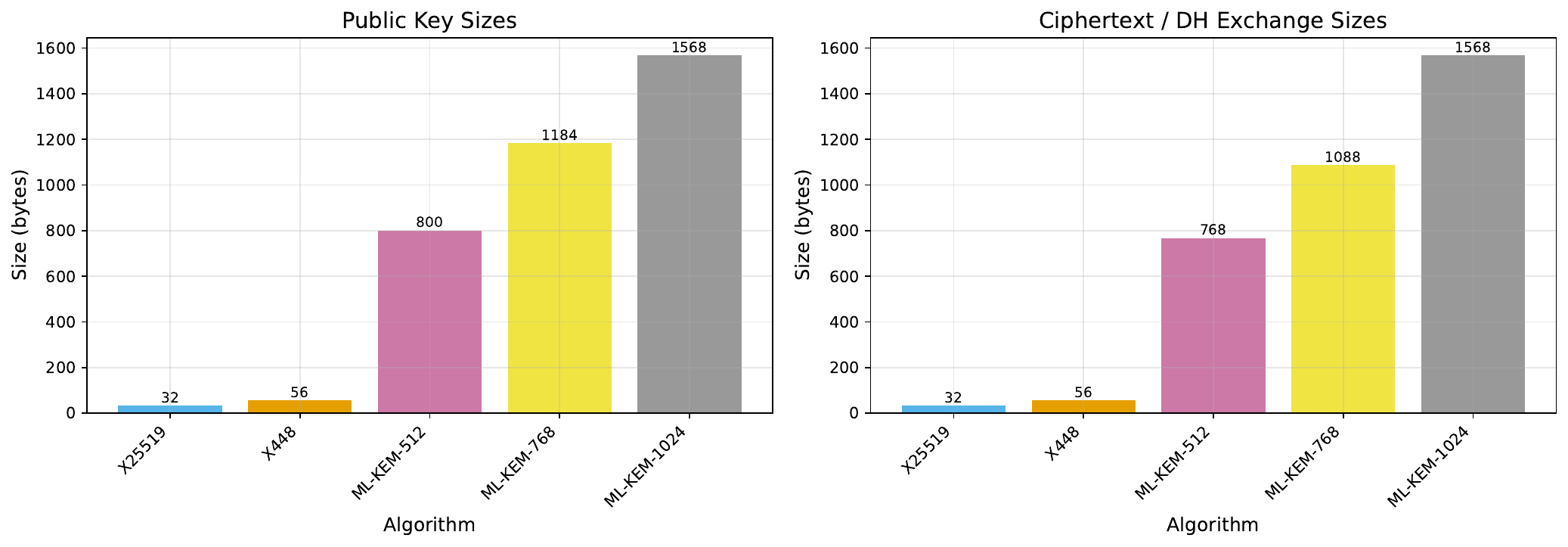}
\caption{Public key and ciphertext sizes for DH and ML-KEM.}
\label{fig:keysizes}
\end{figure}

\begin{figure}[!t]
\centering
\includegraphics[width=\columnwidth,trim=6 6 6 6,clip]{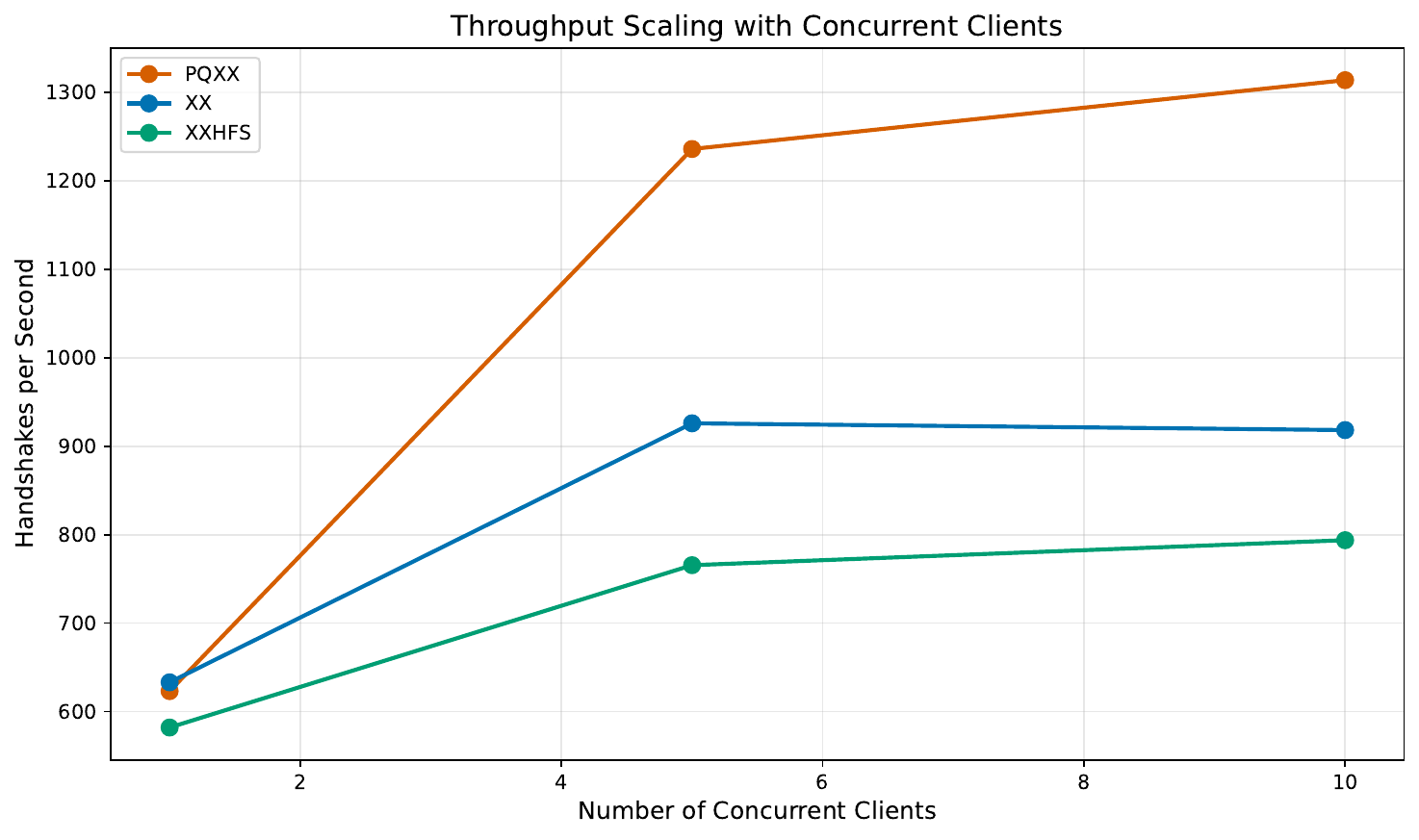}
\caption{Throughput Scaling with Concurrent Clients.}
\label{fig:throughput}
\end{figure}

\subsection{Performance and Benchmarks}

We evaluate \textbf{\textit{NoisePQC++}} across seven performance metrics using Google Benchmark\cite{GoogleBenchmark2026} and ASIO networking library\cite{AsioLibrary} on an Intel i7 (2.30 GHz, 32 GB RAM) MacBook Pro under \textbf{\textit{Clang 18}} compiler with \textit{Release} optimizations. The results we present next are an average over several runs of the benchmarking scripts.

\paragraph{Handshake Latency Comparison}
ML-KEM-768 handshakes are consistently faster than X25519, with overhead ratios of 0.66--0.85$\times$ as shown in Fig.~\ref{fig:latency}. This likely reflects efficient lattice arithmetic and Botan's optimized x86-64 implementation. Among all the interactive patterns, the \textbf{NN} pattern is the fastest in all three modes, while \textbf{IK} and \textbf{KK} are slower due to additional static-key operations. HFS increases cost to 1.25--1.70$\times$ over classical because each handshake performs both ECDH and KEM, trading moderate overhead for defense-in-depth.

\paragraph{CPU Time Analysis}
CPU time closely matches wall-clock time across all modes, with CPU/real ratios above 0.98, indicating computation-bound behavior (Fig.~\ref{fig:cpu}). Classical patterns show the widest spread because of the X448 versus X25519 gap, whereas PQ patterns cluster more tightly across ML-KEM levels. HFS timings are roughly additive, reflecting the combined cost of ECDH and KEM within a single handshake.

\paragraph{Key Generation Performance}
Key generation helps explain the latency results as shown in Fig.~\ref{fig:keygen}. ML-KEM-768 takes 59.1 $\mu$s, nearly identical to X25519 at 57.2 $\mu$s, while ML-KEM-512 is the fastest primitive at 34.9 $\mu$s. Even ML-KEM-1024 remains below 100 $\mu$s at 93.3 $\mu$s, scaling roughly with security level ($512\to768\to1024 \approx 1.0\times\to1.7\times\to2.7\times$). In contrast, X448 is 4.0$\times$ slower than X25519 (227.9 $\mu$s vs. 57.2 $\mu$s), showing that PQ handshakes benefit from both efficient encapsulation and competitive key generation.

\paragraph{Key and Ciphertext Sizes}
The main cost of post-quantum migration is bandwidth, not computation. ML-KEM-768 public keys and ciphertexts are much larger than X25519 (1184 vs. 32 bytes and 1088 vs. 32 bytes), although the HKDF-based key schedule reduces every exchange to the same 32-byte symmetric keys, so downstream symmetric encryption is unchanged. X448 increases key size only modestly (56 vs 32 bytes), making this ECDH$\rightarrow$KEM transition the dominant source of wire overhead, as shown in Fig.~\ref{fig:keysizes}. ML-KEM-1024 is notable in that its public key and ciphertext are both 1568 bytes.

\paragraph{Message Overhead Analysis}
These size increases compound across handshake messages. PQ patterns incur 28.6--46.8$\times$ more wire overhead than classical, with \textbf{KK} highest because its classical baseline is very small as detailed in Fig.~\ref{fig:overhead}. HFS grows less than pure PQ in multi-message patterns (e.g., \textbf{XX}: HFS 14.4$\times$ vs. PQ 36.1$\times$) because it retains compact classical ECDH tokens while adding a KEM operation instead of replacing all ECDH operations. \textbf{NNhfs} is the exception at 33.0$\times$ because classical \textbf{NN} starts from only 80 bytes.

\paragraph{Network Round-Trip Time}
TCP loopback measurements confirm the local benchmark results under transport overhead. \textbf{pqNN} achieves the lowest mean RTT at 425.0 $\mu$s, outperforming classical NN at 516.2 $\mu$s despite larger payloads, indicating that computation rather than payload size dominates in loopback as detailed in Fig.~\ref{fig:rtt}. Variance is higher than in local benchmarks because of OS scheduling, TCP stack latency, and context switching, but P99 latency remains below 2 ms in all modes. \textbf{XXhfs} is the slowest, with 1186 $\mu$s mean RTT, due to three protocol messages and dual ECDH+KEM operations.

\paragraph{Throughput Scaling}
For this benchmark, we only selected the \textbf{XX} pattern to compare the three modes of classical, PQ only, and hybrid. Under concurrent load, \textbf{pqXX} scales best, reaching 1313.9 hs/s at 10 clients versus 918.3 hs/s for \textbf{XX} and 793.8 hs/s for \textbf{XXhfs} as shown in Fig.~\ref{fig:throughput}. This advantage follows from its lower per-handshake CPU cost, which allows more handshakes to complete under contention. \textbf{XXhfs} also shows the highest tail latency under load (P99: 11.1 ms at 10 clients vs. 9.9 ms for \textbf{XX}), reflecting the cost of combined ECDH+KEM operations. All modes improve up to 5 clients, but \textbf{XX} and \textbf{XXhfs} show diminishing gains from 5 to 10 clients, suggesting near-saturation on the 8-core CPU, whereas \textbf{pqXX} still has headroom.

\subsection{Summary of Our Findings}

Overall, the evaluation shows that \textit{\textbf{NoisePQC++}} \textbf{meets its main goals} across coverage, correctness, and performance. It provides broader protocol support than prior implementations, supports practical confidence through testing and instrumentation, and shows that post-quantum and hybrid Noise handshakes remain feasible with manageable overhead. This overhead is modest in practice. The largest ephemeral structure is an ML-KEM key pair ($\sim$2.3 KB for ML-KEM-1024 keys, plus minor overhead), and total handshake state remains only a few kilobytes. C++ RAII ensures prompt release of temporary resources. For side-channel resistance, we rely on Botan's constant-time implementations and clear sensitive material where appropriate. We do not currently implement SEEC from PQNoise\cite{angel2022post}, instead relying on Botan's cryptographic PRNG for secure key generation. These results suggest that \textbf{\textit{NoisePQC++}} is not only a research prototype, but also a practical basis for experimentation and evaluation toward future deployment of quantum-resistant secure channels.

\section{Conclusion and Future Work}
\label{sec:conclusion}

Building on the PQNoise designs and the hybrid-forward-secrecy extension, \textbf{\textit{NoisePQC++}} demonstrates that the Noise Protocol Framework can be extended to support post-quantum and hybrid key exchange while preserving the simplicity and flexibility that makes Noise attractive. By unifying classical, post-quantum, and hybrid handshakes in a single architecture, our work helps bridge the gap between proposed PQ and hybrid Noise designs and a usable software framework. Compared with prior implementations, \textbf{\textit{NoisePQC++}} provides broader pattern coverage, support for multiple ML-KEM parameter sets, and a unified architecture spanning classical ECDH, PQ KEM, and hybrid combinations. The implementation also includes practical tooling for integration, debugging, and experimentation, making it useful both as a library and as a platform for studying post-quantum Noise handshake behavior.

Our evaluation indicates that this broader support does not impose prohibitive cost. Post-quantum and hybrid Noise handshakes remain practical, with manageable latency and computational overhead, while hybrid modes provide a prudent migration path by combining classical and post-quantum assumptions within a single handshake. As NIST-standardized PQC moves toward deployment, \textbf{\textit{NoisePQC++}} provides a concrete step toward quantum-resistant secure channels built on the Noise framework.

Future research will concentrate on five primary directions: (i) extending the experimental evaluation through finer-grained and cross-platform benchmarking; (ii) integrating \textbf{\textit{NoisePQC++}} into practical applications, including prototype implementations for virtual private networks (VPNs) and secure messaging systems\cite{zhang2026toward}; (iii) implementing fallback and negotiation mechanisms to enhance interoperability across heterogeneous deployment environments; (iv) investigating post-quantum digital signature schemes for authentication; and (v) incorporating HQC, the code-based KEM selected by NIST in March 2025 to complement ML-KEM\cite{NIST_IR8545_2025}, and benchmarking it against the ML-KEM parameter sets evaluated here once its standard (HQC-KEM) is finalized. Collectively, these efforts aim to advance \textbf{\textit{NoisePQC++}} toward a fully post-quantum secure channel design that achieves comprehensive coverage of the Noise protocol specification\cite{NoiseSpec34}.

\section*{Code Availability}
The \textit{\textbf{NoisePQC++}} implementation is publicly available as an open-source repository on GitHub~\cite{ahmed2026noisepqcpp}. Reproducibility instructions, Docker image tags, and immutable SHA-256 image digests
are provided in the repository's documentation. The exact version evaluated in this work is archived on Zenodo for long-term access and reproducible citation~\cite{NoisePQCppZenodo}.

\section*{Acknowledgment}

This paper is based upon work supported by the National Science Foundation under award No.~2347249 and US Army under grant No.~W911NF2120076.

\normalsize
\setlength{\itemsep}{0pt}\setlength{\parsep}{0pt} 

\end{document}